\documentclass[10pt,twocolumn,letterpaper]{article}

\usepackage[pagenumbers]{cvpr} % To force page numbers, e.g. for an arXiv version

\definecolor{arxivblue}{rgb}{0.21,0.49,0.74}
\usepackage[pagebackref,breaklinks,colorlinks,allcolors=arxivblue]{hyperref}
\usepackage{bm}
\usepackage{colortbl} % For cell coloring
\usepackage{array}    % For custom column width
\usepackage{graphicx} % For adjusting table spacing
\usepackage{mathabx}
\usepackage{array}
\usepackage{makecell}
\usepackage{wrapfig}
\usepackage{booktabs}
\usepackage{multirow}
\newcommand\blfootnote[1]{%
\begingroup
\renewcommand\thefootnote{}\footnote{#1}%
\addtocounter{footnote}{-1}%
\endgroup
}

\newcommand{\METHOD}{AnimaMimic}

\title{\METHOD: Imitating 3D Animation from Video Priors}

\author{Tianyi Xie$^{1*}$  \quad Yunuo Chen$^{1*}$\quad Yaowei Guo$^{1*}$ \quad Yin Yang$^{2}$ \quad Bolei Zhou$^{1}$ \\
\quad Demetri Terzopoulos$^{1}$ \quad Ying Jiang$^{1}$ \quad Chenfanfu Jiang$^{1}$ \\
$^{1}$ UCLA, $^{2}$ University of Utah
}

\begin{document}
\twocolumn[{%
\renewcommand\twocolumn[1][]{#1}%
\maketitle
\begin{center}
    \vspace{-10pt}
    \centering
    \captionsetup{type=figure}
    \includegraphics[width=\textwidth]{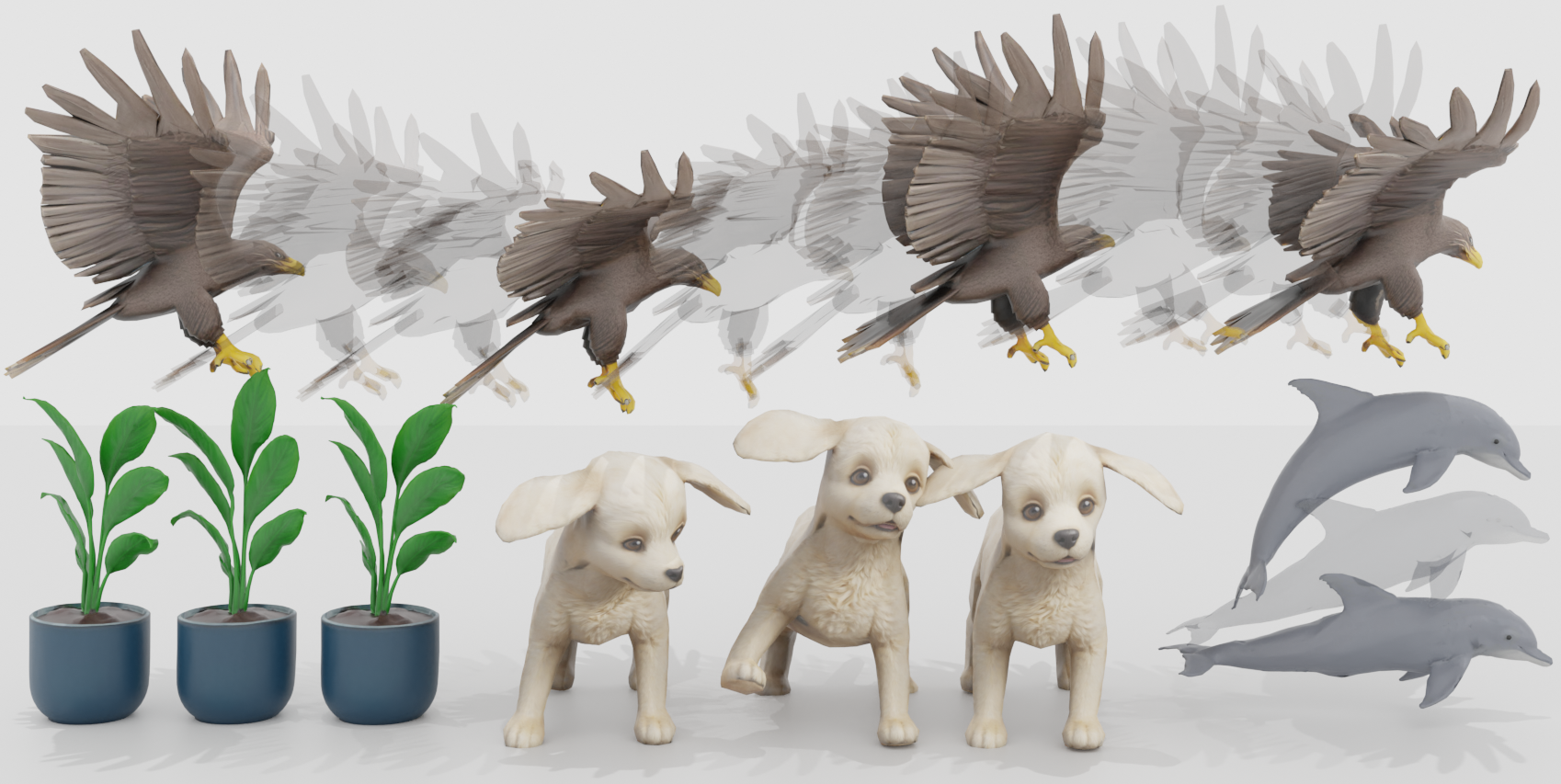}
    \captionof{figure}{\textbf{Animated Creatures.} By optimizing joint articulations and material parameters from videos, our method generates realistic dynamics for objects with diverse geometries.
    }
    \label{fig:teaser}
    \vspace{-5pt}
\end{center}
}]
% \maketitle
% \input{sec/0_abstract}  
 
\begin{abstract}
\blfootnote{* indicates equal contributions.}
Creating realistic 3D animation remains a time-consuming and expertise-dependent process, requiring manual rigging, keyframing, and fine-tuning of complex motions. Meanwhile, video diffusion models have recently demonstrated remarkable motion imagination in 2D, generating dynamic and visually coherent motion from text or image prompts. However, their results lack explicit 3D structure and cannot be directly used for animation or simulation. We present AnimaMimic, a framework that animates static 3D meshes using motion priors learned from video diffusion models. Starting from an input mesh, AnimaMimic synthesizes a monocular animation video, automatically constructs a skeleton with skinning weights, and refines joint parameters through differentiable rendering and video-based supervision. To further enhance realism, we integrate a differentiable simulation module that refines mesh deformation through physically grounded soft-tissue dynamics. Our method bridges the creativity of video diffusion and the structural control of 3D rigged animation, producing physically plausible, temporally coherent, and artist-editable motion sequences that integrate seamlessly into standard animation pipelines. Our project page is at \url{https://xpandora.github.io/AnimaMimic/}.
\end{abstract}

\section{Introduction}
\vspace{-5px}
\label{sec:intro}
Creating expressive 3D animation has long been a cornerstone of computer graphics, powering lifelike digital humans, creatures, and virtual worlds. Despite decades of progress, producing high-quality animation remains an extraordinarily labor-intensive endeavor. Skilled artists must design skeletal rigs, keyframe complex motions, and manually refine subtle deformations to achieve believable results. This process demands both creative intuition and physical insight, making realistic animation difficult to automate or scale.
Meanwhile, recent breakthroughs in video diffusion models have transformed motion synthesis in 2D, generating dynamic and stylistically rich sequences such as human performances \cite{guo2024liveportrait, zhu2024champ, hu2024animate}, cartoon actions \cite{xing2024tooncrafter, li2025tooncomposer}, and cinematic camera movements \cite{yu2025trajectorycrafter, ren2025gen3c}. These models exhibit remarkable motion imagination, yet their outputs reside purely in the image plane—devoid of explicit 3D structure required for downstream applications such as rendering, simulation, or interactive editing.

To bridge this gap, emerging 4D generation approaches attempt to recover temporally coherent 3D motion from video diffusion priors. One line of work \cite{li2025articulated, bahmani2024tc4d, bahmani20244d} applies score-distillation sampling (SDS) \cite{poole2022dreamfusion} to directly optimize a 3D representation (e.g., NeRFs \cite{mildenhall2021nerf} or Gaussian splats \cite{kerbl20233d}) under video diffusion supervision. While conceptually elegant, these methods typically entangle geometry and appearance, leading to slow optimization, limited controllability, and difficulties integrating with standard animation pipelines. Another line \cite{jiang2023consistent4d, wu2024sc4d, xie2024sv4d} reconstructs motion from generated videos via implicit radiance-field or Gaussian representations. However, such representations are not readily compatible with modern computer graphics workflows and are difficult to edit or reuse, making integration into downstream applications cumbersome.

Observing this difficulty, we instead seek to animate an explicit 3D representation, removing the need for geometry reconstruction. Our goal thus becomes: \textit{How can we animate an existing 3D model using the motion priors learned by video diffusion models?} Inspired by traditional animation workflows where artists begin by sketching a skeleton and iteratively refining the motion, we adopt a skeleton-based rigging animation model with a 3D mesh representation. Starting from a static 3D mesh, we first employ an automatic rigging module to extract a skeletal structure that drives mesh deformation. To provide motion guidance, we use differentiable rendering to project the mesh into 2D space and leverage animation videos generated by a video diffusion model for video-space supervision. Through gradient-based optimization, our method aligns the animated mesh poses with these motion cues, producing temporally coherent and editable animations.

% \TODO{The following two paragraphs need to be revised}
% In this work, we take a more practical and interpretable perspective: \textit{How can we animate an existing 3D model using the motion priors learned by video diffusion models?} 
% Inspired by traditional animation workflows, where artists begin by sketching a skeleton and iteratively refining motion, we present \METHOD, a framework that mimics 3D animation from video priors. 

To further enhance dynamic realism, we introduce a physics-aware motion-refinement module that models soft-tissue dynamics. By embedding a differentiable physics-based simulator into the animation loop, our method enforces physical plausibility and temporal smoothness while preserving the creative flexibility of data-driven motion synthesis. The resulting animations exhibit natural, physics-grounded dynamics and strong correspondence to real-world motion. In summary, our main contributions include:
\begin{itemize}
\item A novel framework, \METHOD, that animates 3D meshes using motion priors distilled from video diffusion models.
\item Differentiable rigging and pose optimization that align 3D skeletal motion with video-based cues through differentiable rendering and gradient-based optimization.
\item A physics-based motion-refinement module seamlessly integrated into the optimization loop, enabling physically plausible and temporally coherent results.
\end{itemize}

\section{Related Work}
\label{sec:formatting}

\paragraph{4D Generation}
4D generation, which incorporates the temporal dimension into 3D geometry, has become an active yet challenging area of research. A large number of research works have emerged in recent years. The first line of research focuses on text-to-4D generation 
% \cite{chai2024star, zheng2024unified, singer2023text4d, bah20244dfy, chen2024ct4dconsistenttextto4dgeneration, ling2024alignyourgaussians, zeng2024trans4d, xu2024comp4d, cao2024avatargo, yu20244realphotorealistic4dscene, hong2022avatarclipzeroshottextdrivengeneration, miao2024pla4dpixellevelalignmentstextto4d, deng2025stp4dspatiotemporalpromptconsistentmodeling, dai2025textmesh4dhighqualitytextto4dmesh, mou2024instruct4dto4dediting4d, shao2023control4d} 
\cite{bah20244dfy, xu2024comp4d, deng2025stp4dspatiotemporalpromptconsistentmodeling, dai2025textmesh4dhighqualitytextto4dmesh, mou2024instruct4dto4dediting4d} 
which adopts diffusion-based generative models to produce reference images or videos, and employs SDS to optimize 3D geometry and motion. Another line of research focuses on image-to-4D generation 
% \cite{10.1007/s11263-023-01839-1, yang2024diffusion, pangdisco4d, ren2023dreamgaussian4d, sang2025twosquared4dgeneration2d, li20244k4dgenpanoramic4dgeneration, zhou2025holotimetamingvideodiffusion} 
\cite{10.1007/s11263-023-01839-1, yang2024diffusion, sang2025twosquared4dgeneration2d, li20244k4dgenpanoramic4dgeneration, zhou2025holotimetamingvideodiffusion} 
and video-to-4D generation 
% \cite{guo2024dist4d, park2025zero4dtrainingfree4dvideo, jiang2024consistentd, Yang_2025_ICCV, xie2025sv4ddynamic3dcontent, zhang20244diffusionmultiviewvideodiffusion, jiang20234deditorinteractiveobjectlevelediting, ren2024l4gm, pan2024efficient4d, wang2024vidu4d, zhu2025ar4dautoregressive4dgeneration, huang2025mvtokenflow, zeng2024stag4d, nag2025in24dinbetweeningsingleviewimages, li2025fb4dspatialtemporalcoherentdynamic, wang2025video4dgenenhancingvideo4d, wu2024cat4d, li2024dreammesh4d} 
\cite{wang2024vidu4d, zhu2025ar4dautoregressive4dgeneration, huang2025mvtokenflow, nag2025in24dinbetweeningsingleviewimages, li2024dreammesh4d} 
which generates multiview video sequences for direct supervised learning, while others rely on skinning techniques such as Linear Blend Skinning (LBS) to deform 3D meshes and ensure consistent motion modeling.
Multi-conditional 4D generation 
% \cite{zhao2023animate124, yang2025learningcoherentmatrixizedrepresentation, sun2024eg4d, jiang2024animate3d, lin2024phy124fastphysicsdriven4d, cai2023genren, wu2025animateanymesh, bah2024tc4d, wang2023drivedreamerrealworlddrivenworldmodels, liang2024diffusion4d, yin20234dgen, liu2025free4d, zhou2025coco4dcomprehensivecomplex4d, yuan20244dynamictextto4dgenerationhybrid, zhao2024genxd, fu2024sync4dvideoguidedcontrollable, shao2024human4dit, dreamdrive, wang2024stag-1, wang2024occsora, gao2024magicdrive3d}
\cite{sun2024eg4d, jiang2024animate3d, wu2025animateanymesh, zhao2024genxd, shao2024human4dit} 
aims to enhance the fidelity and coherence of synthesized motion by integrating multiple modalities and diverse training strategies. However, these vast majority of 4D generation frameworks do not directly use high-quality 3D targets for generation due to their scarcity. 

Recently, 3D-to-4D generation has attracted increasing attention by directly taking 3D objects as input and transforming static geometry into dynamic motion. 
% Motion3DGAN \cite{otberdout2022sparse} deforms static facial meshes using GAN-predicted 3D facial displacements. 
HyperDiffusion \cite{erkoç2023hyperdiffusion} learns directly over neural field parameters to synthesize dynamic meshes. 
ElastoGen \cite{feng2024elastogen} models realistic elastic dynamics by using 3D CNN to refine deformation fields from static 3D models. Animate3D \cite{jiang2024animate3d} animates static 3D models using a multi-view video diffusion model that synthesizes multi-view motion sequences. AKD \cite{li2025articulatedkinematicsdistillationvideo} and Puppeteer \cite{song2025puppeteer} all builds a Skeleton-Rigging-Skinning pipeline to generate 4D animations. 
% However, these approaches that optimizes joint parameters using differentiable rendering and applies LBS to deform the mesh surface output quasi-static animation which lacks physically-grounded motion. In contrast, our framework leverages differentiable simulation to model inertia and elasticity for more physically consistent and realistic dynamic behavior.
However, prior LBS-based methods that optimize joints via differentiable rendering yield quasi-static motion lacking physical realism. Our framework instead employs differentiable simulation to capture inertia and elasticity, producing physically consistent, realistic dynamics.

\vspace{-10px}
\paragraph{Skeleton Generation}
Automatic skeleton generation is fundamental to enabling articulated 3D models to deform realistically. Conventional skeleton generation methods primarily rely on geometric analysis on mesh topology and spatial features to infer skeletal structures without the need for learning. Pinocchio \cite{10.1145/1276377.1276467} embeds skeletons into 3D meshes by optimizing medial-surface–based embeddings following a penalty-driven approach. 
% Tagliasacchi et al. \cite{10.1145/1531326.1531377} proposes constructing the curve skeleton of an object by identifying a generalized rotational symmetry axis (ROSA) in oriented point-cloud data. 
% MCS \cite{taglia_sgp12} extracts curve skeletons by contracting meshes through mean curvature flow to reveal medial structures. 
Ju et al. \cite{10.1007/11802914_17} generates topological preserving skeletal graphs by applying iterative morphological thinning to volumetric models. However, conventional skeleton generation methods rely on geometric heuristics and lack semantic understanding and articulation awareness for realistic deformation.

Recent advances in deep learning have transformed skeleton generation from heuristic geometric analysis into a data-driven prediction problem, and a series of related works \cite{RigNet, chu2024humanriglearningautomaticrigging, VINECS, song2025puppeteer} have recently emerged. RigNet \cite{RigNet} predicts both a skeleton structure and corresponding skinning weights directly from a 3D mesh using a deep neural architecture trained on a large dataset. HumanRig \cite{chu2024humanriglearningautomaticrigging} applies a coarse-to-fine structure to predict and refine human's skeletons and skinning weights. 
% VINECS \cite{VINECS} introduces a fully automated pipeline that infers a human's rig and pose-dependent skinning weights from multi-view videos of a human actor. 
Puppeteer \cite{song2025puppeteer} and UniRig \cite{zhang2025modelrigalldiverse} generate skeleton hierarchies and skinning weights using auto-regressive transformers with joint or tree structured tokenization for automatic rigging. 

\vspace{-10px}
\paragraph{Physics-based Animation}
% complimentary dynamic related, computer animations, 
% first paragraph says general works, such as interpolation for character animaiton siggraph(CharacterMixer), Motion2Motion, physics simulation for animation siggraph

% second paragraph describes complimentary dynamic(Rig-Space Physics), Fast Complementary Dynamics paper

% Physics-based animation has been a long-standing research topic in computer graphics and animation, 
Physics-based animation aims to produce realistic motion by modeling underlying physical laws rather than relying solely on artist-defined key frames. Some of these works focus on interpolation for character animation \cite{zhan2024charactermixerrigawareinterpolation3d, eisenberger2021neuromorphunsupervisedshapeinterpolation, zheng2022sdfstyleganimplicitsdfbasedstylegan}, aiming to generate smooth and natural transitions by blending existing 3D characters with different mesh and skeleton structures. NeuroMorph \cite{eisenberger2021neuromorphunsupervisedshapeinterpolation} uses a graph-convolutional architecture to produce smooth shape interpolations and point-to-point correspondences between two 3D shapes. CharacterMixer \cite{zhan2024charactermixerrigawareinterpolation3d} builds a unified skeleton by interpolating bone distance fields to enable interpolation across diverse character. Other works follow the idea of motion transfer across characters \cite{chen2025motion2motioncrosstopologymotiontransfer, li2022iterative, zhang2023semantic}, trying to transfer motions from one character to another while preserving motion style and physical constraints. Zhang et al. \cite{zhang2023semantic} leverages a vision-language model to align motions embeddings, enabling skeleton-aware and semantically consistent motion retargeting. Motion2Motion \cite{chen2025motion2motioncrosstopologymotiontransfer} enables motion transfer across characters by matching and blending motion patches according to sparse joint correspondences. Another line of work focuses on motion generation and simulation \cite{yuan2023physdiff, tevet2022humanmotiondiffusionmodel, zhang2022motiondiffuse, ren2023diffusionmotiongeneratetextguided}, such as PhysDiff \cite{yuan2023physdiff} which incorporates physical constraints into diffusion models to generate physically consistent human motions.

A closely related line of research focus on incorporating complementary dynamics to enhance primary motion with secondary effects such as inertia, elasticity, and soft-body responses \cite{benchekroun2023fast, 10.1145/2601097.2601116, 10.1145/3596711.3596794, 10.1016/j.jvcir.2007.01.005}. Zhang et al. \cite{ZhangCompDynamics2020} proposes optimizing elastodynamic displacements in the subspace orthogonal to a rig’s controllable motions, thereby enriching rig animations with secondary physical effects without interfering with the original rig-driven motion. Building upon \cite{ZhangCompDynamics2020}, Benchekroun et al. \cite{benchekroun2023fast} further accelerate complementary dynamics computation by parameterizing a skinning-based subspace with eigenmodes, enabling efficient, rotation-equivariant, and real-time simulation of secondary motion.

\section{Method}
\vspace{-5px}
\begin{figure*}[t]
    \centering
    % --- Placeholder box for the method overview figure ---
    \includegraphics[width=\textwidth]{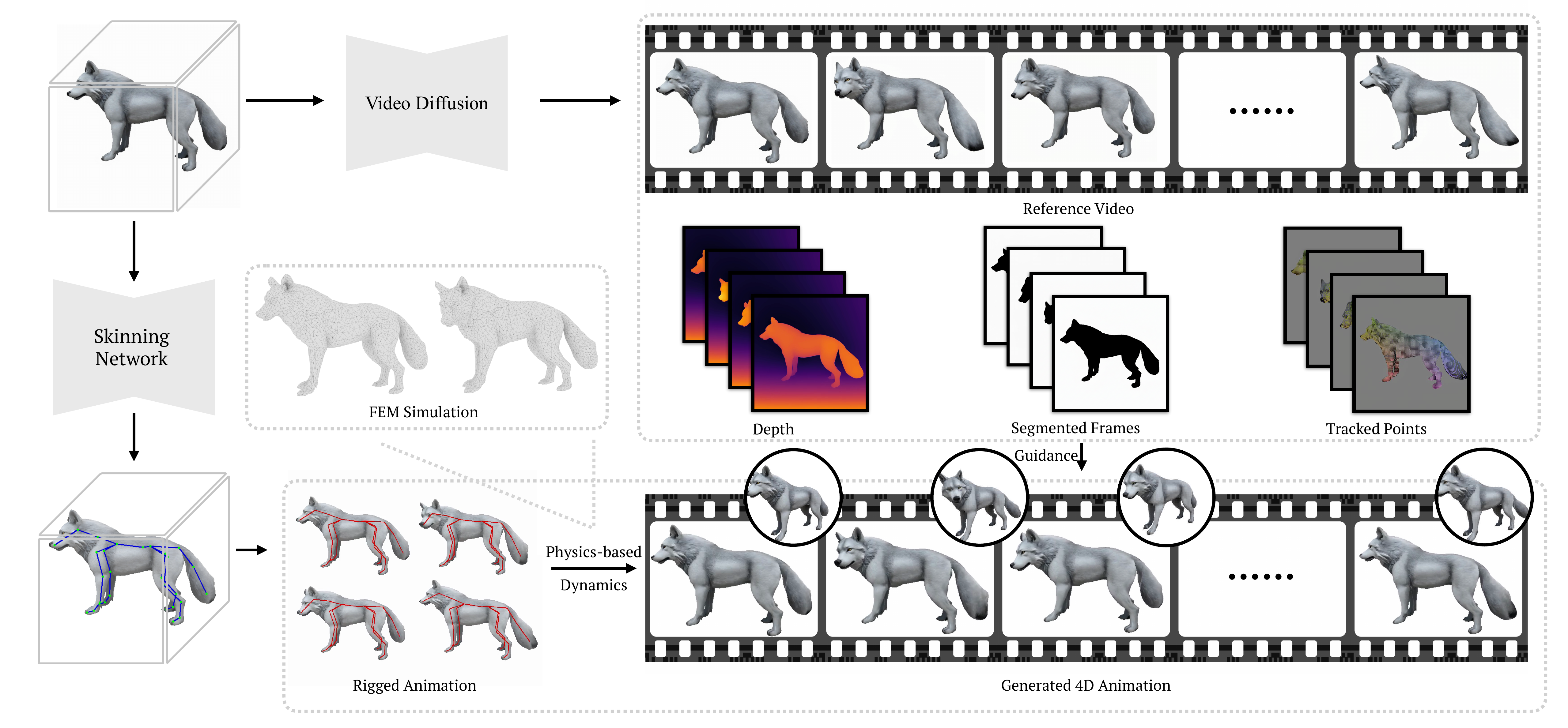}
    \caption{
        \textbf{Pipeline Overview.} From an input 3D mesh, we render a canonical view and use a video diffusion model to generate a monocular motion sequence. We construct a skeleton with skinning weights using a feed-forward rigging model and generate animation by optimizing joint motions through differentiable rendering, tracking, and depth cues. Finally, we refine mesh deformation via differentiable simulation to obtain physically grounded and temporally consistent results. Right circles indicate novel views.
    }
    \vspace{-10px}
    \label{fig:pipeline}
\end{figure*}

We propose \METHOD, a novel framework that transforms a static textured mesh $\mathcal{S}$ into an animated sequence $\{\mathcal{S}^1, \mathcal{S}^2, \mathcal{S}^3, ..., \mathcal{S}^T\}$. Given an input 3D mesh, either artist-created or generated by text/image-to-3D models, our proposed framework first renders the mesh into a canonical view image $I$ and employs a video diffusion model to produce a monocular video sequence $\{I^1, I^2, I^3, ..., I^T\}$ that captures rich motion priors. We then employ a feed-forward rigging network to automatically construct the skeleton with skinning weights, converting the mesh into an animation-ready representation. We further optimize the animation through a two-stage process: the first stage (Sec.~\ref{sec:rigging animation}) aims to optimize joint parameters, including rotations and translations, via differentiable rendering, tracking, and depth supervision to align the mesh motion with video-based cues; the second stage (Sec.~\ref{sec:secondary dynamics}) incorporates differentiable deformable-body simulation to refine the motion with physics-based deformation. This unified pipeline effectively bridges the creative motion imagination of video diffusion models with the structural controllability of 3D rigged mesh animation, producing dynamic, editable, and physically consistent motion sequences. Fig.~\ref{fig:pipeline} provides an overview of our pipeline.

\vspace{-3px}
\subsection{Rigging-based Animation}
\vspace{-3px}
\label{sec:rigging animation}
To make the animation optimization stable and tractable, we represent mesh motion using a low-dimensional rigging formulation rather than directly optimizing all vertex positions. 
Directly manipulating vertex positions is computationally expensive and often leads to unstable or suboptimal results due to the high-dimensional search space. 
Inspired by traditional animation workflows, we model motion through a rigging system and optimize the joint parameters by aligning the animated mesh with a generated reference video.

\subsubsection{Rigging and Skinning}

A rigging system consists of a root joint $J_0$ and a hierarchy of joints $\{J_i\}_{i=1}^K$, each associated with a local transformation parameterized by a rotation $\mathbf{R}_i \in SO(3)$ and a translation $\mathbf{t}_i \in \mathbb{R}^3$. 
The global transformation of each non-root joint is computed via forward kinematics (FK)~\cite{denavit1955kinematic}, which recursively propagates local transformations along the kinematic chain:
\begin{equation}
    \mathbf{T}_i = \mathbf{T}_{\text{parent}(i)} \, [\mathbf{R}_i \,|\, \mathbf{t}_i],
\end{equation}
where $\mathbf{T}_i \in SE(3)$ denotes the global transformation of joint $J_i$. 
This hierarchical formulation ensures coherent motion propagation along the articulated structure.

To deform the surface mesh according to skeletal motion, we adopt Linear Blend Skinning (LBS)~\cite{magnenat1989joint}, a widely used and efficient deformation model in computer animation. 
Each vertex with position $\mathbf{X}_i$ in the rest pose is influenced by a set of nearby joints through predefined skinning weights $\{w_{ik}\}_{k=1}^K$. 
The deformed vertex position $\mathbf{x}_i$ is given by:
\begin{equation}
    \mathbf{x}_i = \sum_{k=1}^{K} w_{ik} \, \mathbf{T}_k \, \mathbf{X}_i, 
    \quad \text{where} \quad \sum_{k} w_{ik} = 1.
\end{equation}
The skinning weights determine the contribution of each joint to vertex deformation, producing smooth transitions between rigid parts while preserving local structure.

We use UniRig~\cite{zhang2025one}, a feed-forward network trained on large-scale rigged 3D models, to automatically predict both the skeletal hierarchy and skinning weights for arbitrary 3D meshes. 
We then parameterize the pose of the 3D model using a set of transformation parameters $\bm{\theta} = \{ (\mathbf{r}_i, \mathbf{t}_i) \}_{i=0}^K$,
where $\mathbf{r}_i$ is the 6D rotation representation~\cite{zhou2019continuity} and $\mathbf{t}_i$ is the translation vector for joint $J_i$. Each 6D rotation vector $\mathbf{r}_i$ can mapped to a valid rotation matrix via the continuous 6D-to-$SO(3)$ conversion function $\mathbf{R}_i = \mathbf{R}(\mathbf{r}_i)$.

\subsubsection{Joint Parameter Optimization}
We aim to leverage the motion priors embedded in video foundation models to automatically animate a 3D mesh. 
To achieve this, we first render a reference image $I$ of the input 3D model from a user-specified camera view. 
Given either a user-provided or VLM-generated text prompt, the reference image is then fed into the video diffusion model to synthesize a sequence of frames
$\{I^t\}_{t=1}^{T} = \{I^1, I^2, \ldots, I^T\}$ as the image-space motion reference. 

We then formulate an optimization problem to lift this 2D motion into 3D space by solving for the joint transformation parameters across all frames. Formally, we optimize:
\begin{equation}
    \min L(\{\bm{\theta}^t\}_{t=1}^{T}; \mathcal{S}, \{I^t\}_{t=1}^{T}, \{J_i\}_{i=0}^{K}, w),
\end{equation}
where $\bm{\theta}^t = \{(\mathbf{r}_i^t, \mathbf{t}_i^t)\}_{i=0}^{K}$ denotes the set of joint transformation parameters to be optimized at frame $t$. The overall objective function $L$ comprises several components:
\begin{equation}
    \begin{aligned}
            L = &\lambda_{\text{rgb}}L_{\text{rgb}} + \lambda_{\text{mask}}L_{\text{mask}} + \lambda_{\text{track}}L_{\text{track}} \\
       & + \lambda_{\text{depth}}L_{\text{depth}} + \lambda_{\text{smooth}}L_{\text{smooth}} + \lambda_{\text{reg}}L_{\text{reg}}
    \end{aligned}
\end{equation}

We detail the loss components below.
% The following details the components of this loss function.

\vspace{-10px}
\paragraph{Differentiable Rendering Loss.} 
The objective of this optimization is to align the rendered animation sequence with the reference video. To obtain accurate spatial supervision, we extract the ground-truth mask sequence $\{M^1, M^2, \ldots, M^T\}$ using SAM2~\cite{ravi2024sam}. We then employ differentiable SoftPhong and silhouette rendering from PyTorch3D~\cite{ravi2020accelerating} to backpropagate gradients through the rendering process to optimize the joint parameters. 
We define two differentiable rendering losses:
\begin{equation}
    L_{\text{rgb}} = \frac{1}{T} \sum_{t=1}^{T} 
    \big\| \mathcal{R}^I(\mathcal{S}, \bm{\theta}^t) - I^t \big\|,
\end{equation}
\begin{equation}
    L_{\text{mask}} = \frac{1}{T} \sum_{t=1}^{T} 
    \big\| \mathcal{R}^M(\mathcal{S}, \bm{\theta}^t) - M^t \big\|,
\end{equation}
where $\mathcal{R}^I(\cdot)$ and $\mathcal{R}^M(\cdot)$ denote the differentiable SoftPhong and silhouette renderers, respectively.

\vspace{-10px}
\paragraph{Tracking Loss.}
The videos generated by video diffusion models can exhibit hallucinated regions and inconsistent appearances caused by lighting variations and changing shadows. To provide more stable image-space supervision, we sample a set of the pixels $\mathcal{P}$ from the foreground region of the reference image $I$, where $\mathcal{P}_i$ denotes the coordinate of $i$-th sampled pixel. Each pixel is then unprojected onto the 3D mesh to identify the corresponding triangle that contains it, and we record its barycentric weights $\beta_i$. We then employ AllTracker~\cite{harley2025alltracker} to track the 2D trajectories of these pixels across the generated video, yielding the tracked pixel sets $\{\mathcal{P}^1, \mathcal{P}^2, ..., \mathcal{P}^T\}$. The tracking loss is defined as:
\begin{equation}
    L_{\text{track}} = \frac{1}{T} \sum_{t=1}^T \big\| \mathcal{B}(\mathcal{S}^t, \bm{\beta}) - \mathcal{P}^t \big\|,
\end{equation}
where $\mathcal{B}(\cdot)$ denotes the barycentric interpolation and the camera projection function that maps 3D points on the mesh $\mathcal{S}^t$ to the corresponding 2D image-space coordinates.

\vspace{-10px}
\paragraph{Depth Loss.}
While differentiable rendering and pixel tracking provide effective image-space supervision, 
they offer limited constraint along the depth dimension. 
To address this, we incorporate additional depth supervision using VGGT~\cite{wang2025vggt}, 
which predicts per-frame depth maps from the generated video. 
For each tracked pixel, we query its corresponding depth value, obtaining a sequence of tracked-point depths 
$\{\mathcal{D}^1, \mathcal{D}^2, \ldots, \mathcal{D}^T\}$. We then define the normalized depth loss as:
\begin{equation}
    L_{\text{depth}} = \frac{1}{T} 
    \sum_{t=1}^{T} 
    \big\| \mathcal{Z}(\mathcal{S}^t, \bm{\beta}) - \mathcal{N}(\mathcal{D}^t) \big\|,
\end{equation}
where $\mathcal{Z}(\cdot)$ is similar to $\mathcal{B}(\cdot)$ but outputs the depth values of the corresponding 3D points in the camera space. A normalization operator $\mathcal{N}(\cdot)$ is applied to mitigate depth-scale ambiguity across frames.
% To mitigate depth-scale ambiguity, we apply a normalization operator $\mathcal{N}(\cdot)$ defined as:
% \begin{equation}
%     \mathcal{N}(\mathcal{D}) = 
%     \frac{\mathcal{D} - \operatorname{median}(\mathcal{D})}
%     {\operatorname{mean}(|\mathcal{D} - \operatorname{median}(\mathcal{D})|)}.
% \end{equation}
% We then define the normalized depth loss as:
% \begin{equation}
%     L_{\text{depth}} = \frac{1}{T} 
%     \sum_{t=1}^{T} 
%     \big\| \mathcal{Z}(\mathcal{S}^t, \bm{\beta}) - \mathcal{N}(\mathcal{D}^t) \big\|,
% \end{equation}
% where $\mathcal{Z}(\cdot)$ is similar to $\mathcal{B}(\cdot)$ but outputs the depth values of the corresponding 3D points in the camera space. 

\vspace{-10px}
\paragraph{Additional Loss.}
% Specifically, we consider both first- and second-order vertex position differences across consecutive frames to ensure gradual transitions:
% \begin{equation}
% \begin{aligned}
%     L_{\text{smooth}} 
%     &= \frac{1}{T-1} \sum_{t=1}^{T-1} \big\| \mathbf{x}^{t+1} - \mathbf{x}^{t} \big\|_2 \\
%     &\quad + \frac{1}{T-2} \sum_{t=2}^{T-1} \big\| \mathbf{x}^{t+1} + \mathbf{x}^{t-1} - 2\mathbf{x}^{t} \big\|_2.
% \end{aligned}
% \end{equation}

To encourage temporal coherence in the reconstructed animated mesh sequence, we introduce a smoothness loss $L_{\text{smooth}}$ that penalizes abrupt motion changes across time. This loss incorporates both first- and second-order vertex differences across consecutive frames to enforce stable and gradual transitions.

We further apply a regularization term to prevent excessive local transformations for non-root joints:
\begin{equation}
    L_{\text{reg}} = \frac{1}{T} \sum_{t=1}^{T} 
    \big\| \bm{\theta}^t - 
    \operatorname{clip}(\bm{\theta}^t, -\hat{\bm{\theta}}, \hat{\bm{\theta}}) \big\|_2,
\end{equation}
where $\hat{\bm{\theta}}$ is the threshold value for rotation and translation magnitudes, and $\operatorname{clip}(\cdot)$ limits each transformation within the allowed range. 

% The overall optimization objective for the rigging-based animation is the weighted combination of all loss terms:
% \begin{equation}
% \begin{aligned}
%     L = &\lambda_{\text{rgb}}L_{\text{rgb}} + \lambda_{\text{mask}}L_{\text{mask}} + \lambda_{\text{track}}L_{\text{track}} \\
%        & + \lambda_{\text{depth}}L_{\text{depth}} + \lambda_{\text{smooth}}L_{\text{smooth}} + \lambda_{\text{reg}}L_{\text{reg}}
% \end{aligned}
% \end{equation}

\vspace{-3px}
\subsection{Physics-based Dynamics}
\label{sec:secondary dynamics}
While the rigging-based animation offers a simple and compact parameterization for capturing motion sequences, it can suffer from surface artifacts caused by imperfect skinning weights predicted by automatic rigging networks. Moreover, the LBS-based deformation model is inherently quasistatic and therefore unable to capture dynamical responses such as inertia effects and elastic response. To overcome these limitations, we further refine the animation sequence using a differentiable physics-based soft-tissue simulation.
\subsubsection{Differentiable Elastic Simulation}
We model the motion as the dynamics of a continuum deformable body. 
Formally, the deformation is represented by a time-dependent map 
$\bm{x} = \phi(\bm{X}, t)$ that transforms material coordinates $\bm{X} \in \Omega^0$ in the undeformed rest space to world coordinates $\bm{x} \in \Omega^t$ in the deformed configuration at time $t$. 
For a discretized tetrahedral mesh, the deformation map of the $i$-th element can be expressed as
$
\phi_i(\bm{X}, t) = \bm{F}_i \bm{X} + \bm{b}_i,
$
where $\bm{F}_i \in \mathbb{R}^{3\times3}$ denotes the deformation gradient and $\bm{b}_i \in \mathbb{R}^3$ is the translation vector. 
The physics-based simulation seeks to solve the motion governed by Newton’s second law:
\begin{equation}
\label{eq:motion equation}
    \frac{d^2 \bm{x}}{dt^2} = \bm{M}^{-1}\bm{f}(\bm{x}),
\end{equation}
where $\bm{M}$ is the mass matrix and $\bm{f}$ represents the internal and external forces acting on the system.

To numerically integrate the motion equation~\eqref{eq:motion equation}, 
we employ the finite element method (FEM) and use the backward Euler scheme for numerical stability, yielding:
\begin{equation}
\begin{aligned}
    \bm{x}^{n+1} &= \bm{x}^n + \Delta t\, \bm{v}^{n+1}, \\
    \bm{v}^{n+1} &= \bm{v}^n + \Delta t\, \bm{M}^{-1}\bm{f}(\bm{x}^{n+1}),
\end{aligned}
\end{equation}
where $\bm{x}^n, \bm{v}^n$ and $\bm{x}^{n+1}, \bm{v}^{n+1}$ denote the positions and velocities at time steps $t^n$ and $t^{n+1}$, respectively. 
This nonlinear system can be formulated as an optimization-based time integration problem~\cite{gast2015optimization} to compute the next state:
\begin{equation}
\label{eq:optim}
    \bm{x}^{n+1} = 
    \arg\min_{\bm{x}} 
    \frac{1}{2} \|\bm{x} - \tilde{\bm{x}}\|^2_{\bm{M}} 
    + \Psi(\bm{x}),
\end{equation}
where $\tilde{\bm{x}} = \bm{x}^n + \bm{v}^n\Delta t + \bm{g} \Delta t^2$ is the predicted position, and $\Psi(\bm{x})$ denotes the total potential energy, including elastic and gravitational terms, such that 
$\bm{f} = -\frac{\partial \Psi}{\partial \bm{x}}$. 
We adopt the Fixed Corotated model~\cite{stomakhin2012energetically} to define the elasticity energy and solve the optimization using Newton’s method with line search for robust and stable convergence.

To enable end-to-end optimization, we follow \cite{li2025dress} to achieve differentiability in simulation, leveraging both the adjoint method and automatic differentiation to backpropagate gradients through the time integration process (Eq.~\ref{eq:optim}). This allows the simulation to remain fully differentiable with respect to $\bm{x}^n$ and the material parameters.
% Details about the differentiability and gradient derivation are provided in the supplementary materials.

\subsubsection{Material Parameter Optimization}
To integrate the differentiable elastic simulation into the animation optimization, 
we first convert the surface triangle mesh $\mathcal{S}$ into a volumetric tetrahedral mesh $\mathcal{S}^{\text{tet}}$ 
using TetWild~\cite{hu2018tetrahedral}, making it suitable for finite element simulation. 
To stabilize training, for each joint $J_i$, we locate its nearest tetrahedron in $\mathcal{S}^{\text{tet}}$ 
and assign the corresponding global transformation $\bm{T}_i$ as the driving boundary condition for the simulation. 
Since TetWild does not guarantee that $\mathcal{S}$ and $\mathcal{S}^{\text{tet}}$ share identical surface topology, 
for each vertex on $\mathcal{S}$ we find its closest tetrahedron in $\mathcal{S}^{\text{tet}}$ and compute its barycentric weights, 
allowing the simulated deformation of $\mathcal{S}^{\text{tet}}$ to drive the motion of $\mathcal{S}$.

To model the elastic stiffness of the body, we consider the Young’s modulus $E_i$ of each tetrahedral element $\tau_i$ and optimize it so that the simulated deformation matches the reference image sequence $\{I^t\}_{t=1}^{T}$. 
This can be formulated as the following optimization problem:
\begin{equation}
    \min_{\{E_i\}_{i=1}^{N}} 
    L(\{E_i\}_{i=1}^{N}; \mathcal{S}^{\text{tet}}, \{I^t\}_{t=1}^{T}, \{\bm{T}_i\}_{i=0}^{K}).
\end{equation}
Here we use the same optimization objective $L$ as in Sec.~\ref{sec:rigging animation}. $\bm{T}_i$ are the joint transformations serving as boundary conditions. 

Directly optimizing $E_i$ for every tetrahedral element can be inefficient and may lead to suboptimal results due to the high degrees of freedom.
To accelerate convergence and enhance stability, we first cluster the tetrahedral elements based on their spatial proximity to each joint, assigning all elements within the same cluster a shared Young’s modulus during optimization.
We then progressively subdivide these clusters to allow finer-grained updates, forming a coarse-to-fine optimization strategy. With the optimized material parameters, the simulation produces a physically realistic animation sequence $\{\mathcal{S}^t\}^T_{t=1}$.

\section{Experiment}

\begin{figure*}[t]
    \centering
    % --- Placeholder box for the method overview figure ---
    \includegraphics[width=\textwidth]{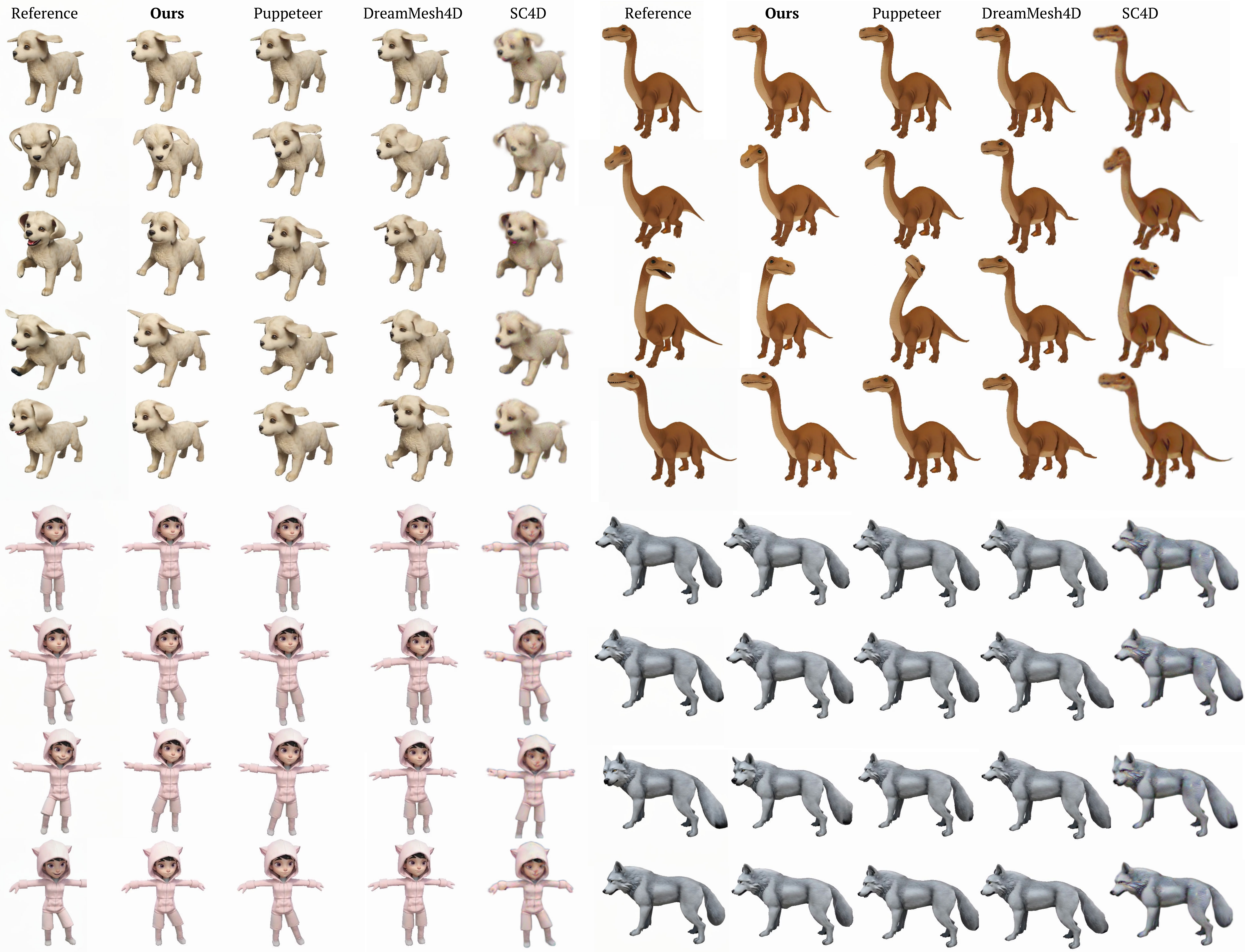}
    \caption{
        \textbf{Qualitative Comparison.} In comparison with SC4D~\cite{wu2024sc4d}, DreamMesh4D~\cite{li2024dreammesh4d}, and Puppeteer~\cite{song2025puppeteer}, our method yields more coherent motion trajectories and more accurately reflects the dynamics present in the reference videos.
    }
    \label{fig:quali-comp}
    \vspace{-10pt}
\end{figure*}

% In this section, we conduct a comprehensive evaluation of our method and compare it against state-of-the-art optimization-based 4D generation approaches.

\paragraph{Datasets}
To evaluate the effectiveness and generality of our framework, we curate a diverse collection of textured 3D models from multiple sources, including public datasets~\cite{RigNet} and commercially available text/image-to-3D generative tools~\cite{meshyai2025, tripo3dstudio2025}. The resulting dataset spans a broad range of object categories, material appearances, and geometric complexities, covering both characters and various animal species.

\vspace{-10px}
\paragraph{Implementation Details}
To generate the reference animation videos, we first render a single frame of the input 3D model and query an LLM to produce a descriptive motion prompt. This prompt, together with the rendered frame, is then fed into the Kling's image-to-video model~\cite{klingai2025} to synthesize the reference animation sequences. For differentiable deformable-body simulation, we implement our physics module using Warp~\cite{warp2022}, which supports AutoDiff operators and is therefore well-suited for differentiable simulation. For both optimization stages, we use the Adam optimizer with a learning rate of $10^{-3}$ to update all parameters. In the simulation-based optimization, we additionally optimize the logarithm of Young’s modulus to improve stability and convergence.
\begin{figure*}[t]
    \centering
    % --- Placeholder box for the method overview figure ---
    \includegraphics[width=\textwidth]{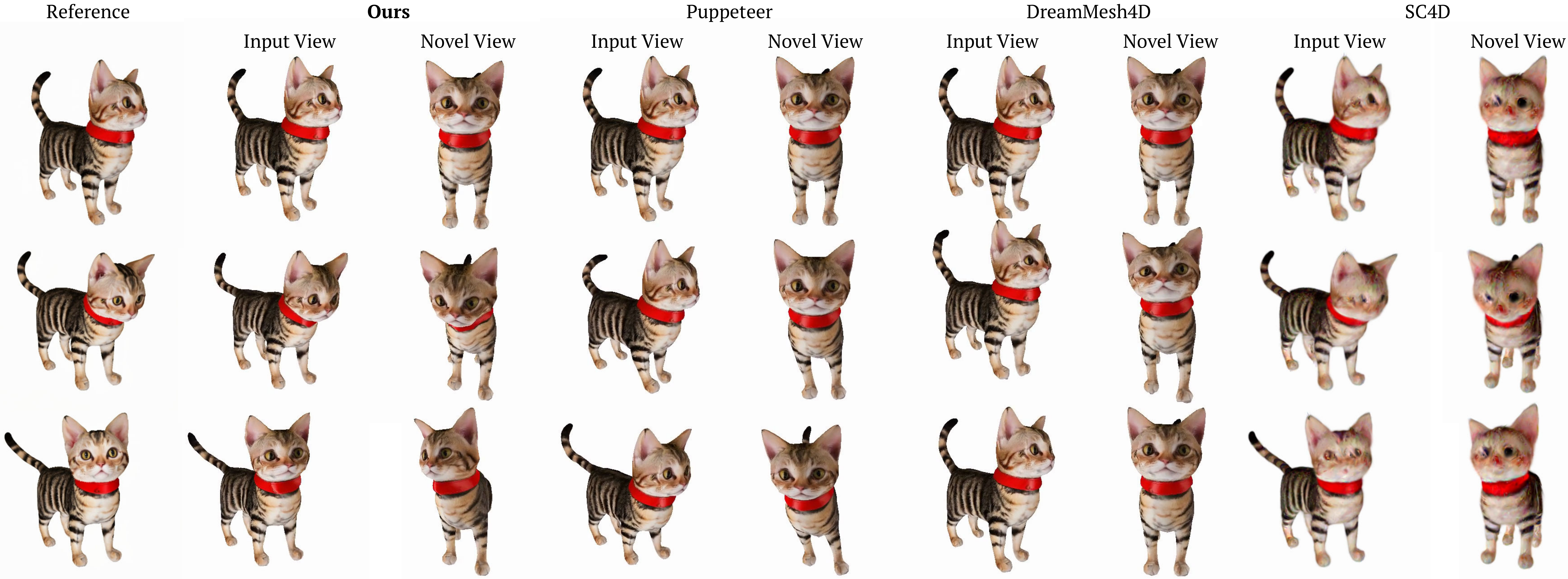}
    \caption{
        \textbf{Novel View Synthesis.} Our method closely aligns with the reference video in the input view and produces coherent novel views, whereas the baseline methods deviate from the ground truth.
    }
    \vspace{-10px}
    \label{fig:quali-novel-view}
\end{figure*}

\vspace{-10px}
\paragraph{Baselines}
We compare our method with state-of-the-art 4D motion generation approaches, including 
SC4D~\cite{wu2024sc4d}, DreamMesh4D~\cite{li2024dreammesh4d}, and Puppeteer~\cite{song2025puppeteer}. 
DreamMesh4D and Puppeteer adopt mesh-based representations, while SC4D relies on Gaussian Splatting. 
Since the original DreamMesh4D requires estimating a 3D model from a single-view image, we use the 
ground-truth mesh as input to ensure a fair comparison. 

% For fairness comparison, we provide the ground-truth mesh for DreamMesh4D, skipping the process of estimating 3d models from single view input. 
% For fair evaluation, we first use recent video generation models\TODO{}~\cite{} to synthesize a motion sequence by rendering the input 3D mesh from a front-view camera and providing the first frame as the initialization. The generated video then serves as input to video-to-4D motion reconstruction pipelines. We evaluate three representative baselines—Puppeteer~\cite{song2025puppeteer}, SC4D~\cite{wu2024sc4d}, and DreamMesh4D~\cite{li2024dreammesh4d}—each of which reconstructs a dynamic 4D sequence from the synthesized video. 
% – A rig-and-animate pipeline that predicts skeletons and skinning weights and optimizes joint rotations using video guidance. Motion is restricted to LBS articulation and cannot model non-rigid surface deformation or generate diverse controllable motions.

% SC4D \cite{wu2024sc4d} – A video-to-4D reconstruction method that decouples motion and appearance via sparse control points and dynamic Gaussians. It reproduces motions observed in the input video but cannot synthesize new, controllable 4D motions for arbitrary 3D objects.

% DreamMesh4D \cite{li2024dreammesh4d} – A video-driven mesh deformation baseline that fits dynamic Gaussians and meshes to a monocular video sequence. Although effective for reconstructing seen motion, it does not support motion control and cannot animate a static 3D asset without a driving reference video

\begin{table}[t]
\centering
\caption{\textbf{Quantitative Comparisons.} We report SSIM and LPIPS to measure the similarity between the generated 4D motion between the referece video. VBAQ, VBOC and VBIQ represents scores for aesthetic quality, overall consistency and  imaging quality, as measured by VBench~\cite{huang2024vbench}.\vspace{-5px}} 
\setlength\tabcolsep{1.8pt}
\label{tab:quantitative comparison}
\small{
\begin{tabular}{p{1.0in}cccccc} 
\hline
\textbf{Methods} & SSIM $\uparrow$ & LPIPS$\downarrow$ & VBAQ$\uparrow$  & VBOC$\uparrow$ & VBIQ$\uparrow$ \\ \hline  %physics score, text score, dynamic score, user consistency score, user preference score
SC4D~\cite{wu2024sc4d}& \cellcolor{Apricot}{0.9403} & 0.0924 & 0.541 & 0.174 & 0.392 \\ 
DreamMesh4D~\cite{li2024dreammesh4d}& 0.8662 & 0.1482 & 0.543 & 0.175 & 0.550 \\ 
Puppeteer~\cite{song2025puppeteer}& 0.9023 & 0.1097 & 0.572 & \cellcolor{Apricot}{0.176} & \cellcolor{Apricot}{0.632} &  \\ 
\hline
Ours & 0.9318 & \cellcolor{Apricot}{0.0849} & \cellcolor{Apricot}{0.581} & \cellcolor{Apricot}{0.176} & 0.606\\ 
\hline
\end{tabular}
\vspace{-5px}
}
\end{table}

\begin{table}[t]
\centering
\caption{\textbf{User Study.} We show the user preference for our method over the baseline methods in terms of  visual quality (VQ), temporal consistency (TC), motion plausibility (MP), and overall feeling. \vspace{-5px}} 
\setlength\tabcolsep{6.0pt}
\label{tab:user study}
\small{
\begin{tabular}{p{1.0in}cccccc} 
\hline
\textbf{Methods} & VQ & TC & MP  & Overall \\ \hline  %physics score, text score, dynamic score, user consistency score, user preference score
SC4D~\cite{wu2024sc4d}& 96\% & 93\% & 92\% & 91\% \\ 
DreamMesh4D~\cite{li2024dreammesh4d}& 88\% & 91\% & 95\% & 91\% \\ 
Puppeteer~\cite{song2025puppeteer}& 74\% & 78\% & 69\% & 71\% \\ 
\hline
\end{tabular}
}
\end{table}

\vspace{-5px}
\subsection{Quantitative Evaluation}
To quantitatively evaluate the generated results, we first measure how well the rendered video matches the input video. We compute SSIM~\cite{1284395}, LPIPS~\cite{zhang2018unreasonable}, and VBench~\cite{huang2024vbench}, which capture overall consistency (VBOC), aesthetic quality (VBAQ) and image quality (VBIQ). For each method, we generate 20 motion sequences and adopt a multi-view rendering protocol: in addition to the input camera view, we render two novel viewpoints by rotating the camera by $\pm 45^\circ$, enabling consistent multi-view comparison across all methods. Additionally, following \cite{geyer2023tokenflow}, we conduct a user study using a two-alternative forced choice (2AFC) protocol, where participants are asked to choose the preferred video based on physical plausibility and overall satisfaction.

The quantitative results in Tab.~\ref{tab:quantitative comparison} show that our method outperforms all baseline approaches in most metrics, including LPIPS, overall temporal consistency, and esthetic quality, demonstrating the superior fidelity of our generated 4D motions. As illustrated in Fig.~\ref{fig:quali-comp}, SC4D tends to produce coarse but roughly accurate motions, achieving high SSIM scores due to preserving low-frequency structure. However, it often generates incorrect textures and distorted geometry, resulting in poor temporal consistency and image quality scores. Puppeteer synthesizes visually sharp and high-quality frames, yet the generated motions are frequently unrealistic, leading to actions that are physically impossible and ultimately yielding lower LPIPS scores. DreamMesh4D maintains closer alignment with the input object’s appearance but is limited to small, near-rigid motions. The user preference results, presented in Tab. 2, show that our proposed method consistently outperforms the baselines across all evaluation criteria. Compared to baselines, users consistently prefer our method for its superior motion plausibility, visual quality, temporal consistency, and overall experience.

% \begin{table*}[t]
% \centering
% \footnotesize
% \renewcommand{\arraystretch}{1.2}
% \setlength{\tabcolsep}{4pt}
% % \resizebox{\textwidth}{!}{
% \caption{\textbf{Quantitative Comparison.} We evaluate overall average SSIM, LPIPS, and FD across all painting styles. Higher SSIM and lower LPIPS/FD indicate better reconstruction fidelity. %\ff{Also color the second place?}
% }
% \begin{tabular}{|l|cc|cccc|}
% \hline
% \multirow{2}{*}{\textbf{Method}} &
% \multicolumn{2}{c|}{\textbf{AAAA}} &
% \multicolumn{4}{c|}{\textbf{VBench}} \\
% \cline{2-7}
%  & \textbf{SSIM}$\uparrow$ &\textbf{LPIPS}$\downarrow$ & \textbf{Aesthetic Quality}$\uparrow $
%  & \textbf{Overall Consistency}$\uparrow$ & \textbf{Motion Smoothness}$\uparrow$ & \textbf{Temporal Flickering}$\uparrow$
%  \\
% \hline
% DreamMesh4D  & 0.8662 & 0.1482        & 0.5432 & 0.1752 & 0.9918 & 0.9907 \\
% \hline
% SC4D  & \textbf{0.9403} & 0.0924          & 0.5411 & 0.1746 & \textbf{0.9957} & \textbf{0.9957} \\
% \hline
% Puppeteer  & 0.9023 & 0.1097          & 0.5722 & 0.1762 & {0.9942} & 0.9933 \\
% \hline
% \textbf{Ours} & {0.9318} & \textbf{0.0849}             & \textbf{0.5804} & \textbf{0.1763} & {0.9942} & {0.9935} \\

% \hline
% \end{tabular}
% % }
% \label{tab:overall}
% \end{table*}

\subsection{Qualitative Evaluation}

\paragraph{Motion Reconstruction}
We present qualitative comparisons with baseline methods in Fig.~\ref{fig:quali-comp}. SC4D frequently produces unclear or unstable motion and, due to its point-based representation, struggles to maintain consistent topology, often resulting in holes or missing regions (e.g., the missing portion of the dog’s tail). DreamMesh4D tends to exhibit distorted geometry around articulated areas such as limbs and joints, due to the large number of independent control nodes to be optimize. Puppeteer generates temporally stable results, but its achievable motion range is noticeably limited. For instance, when the dinosaur turns its neck, it produces only a small upward head tilt instead of a full rotational movement. In contrast, our methods produces physically plausible and visually coherent 4D motion while faithfully preserving the object’s geometry and appearance.

\vspace{-10px}
\paragraph{Novel-view Synthesis} 
Benefiting from our rigged generation framework, our method can synthesize novel-view renderings of the generated motions with strong cross-view consistency. As shown in Fig.~\ref{fig:quali-novel-view}, Puppeteer struggles to maintain alignment between motion and viewpoint. DreamMesh4D better preserves view–motion alignment but exhibits only limited motion range. For instance, the cat remains in nearly the same pose throughout the sequence. SC4D continues to suffer from low-quality textures, and when the cat turns its head, regions previously unseen become blurry. In contrast, our method generates coherent motions with consistent, high-quality appearance across novel views.

\subsection{Ablation Study}

\begin{figure}[h]
    \centering
    \includegraphics[width=\columnwidth]{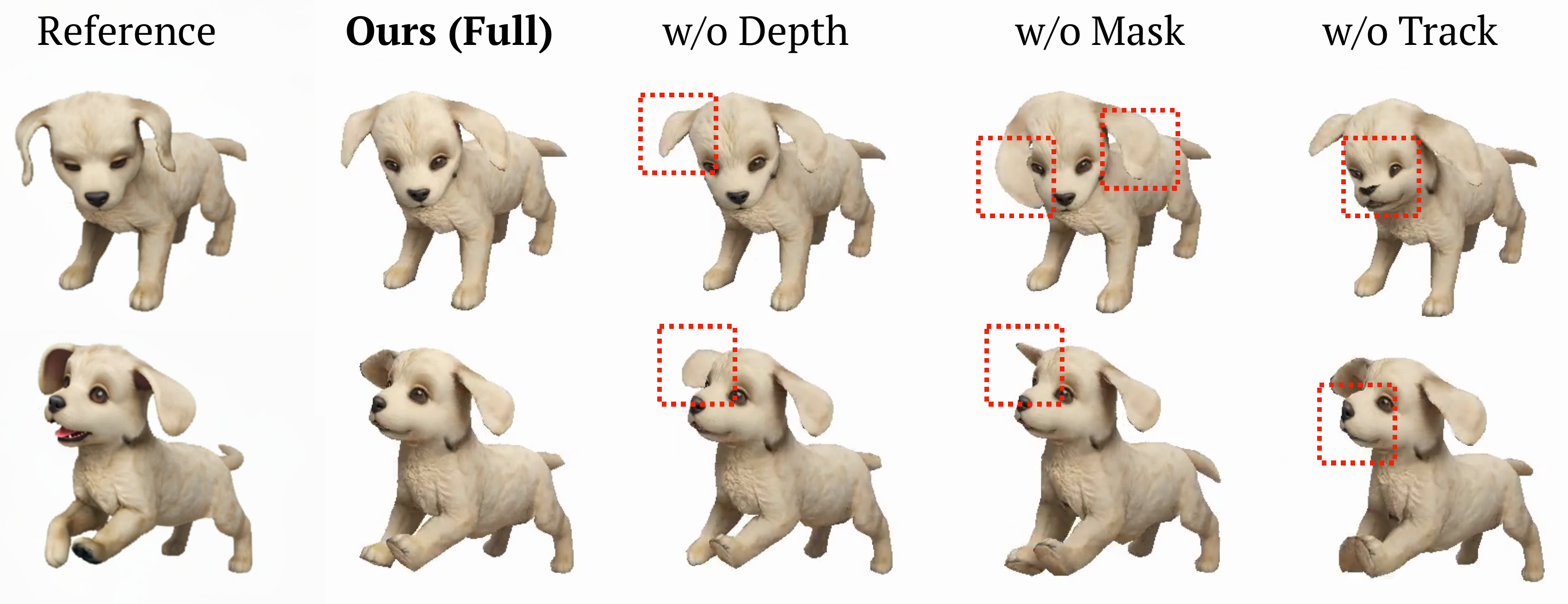}
    \caption{
        \textbf{Ablation Studies.} We conduct ablation studies on the proposed loss terms during optimization. Incorporating these terms leads to more plausible motion and enables the reconstructed dynamics to more faithfully adhere to the input video.
    }
    \label{fig:ablation-loss}
    \vspace{-10pt}
\end{figure}

\paragraph{Loss Components}
To evaluate the efficacy of the proposed loss components, we conduct ablation studies on the loss terms (see Fig.~\ref{fig:ablation-loss}), including depth, mask, and point tracking. We observe that dropping any of these loss terms leads to distorted parts and deviations from the reference motion. Our full model exhibits consistent shape and motion, aligning well with the reference video.

\begin{figure}[h]
    \centering
    \includegraphics[width=\columnwidth]{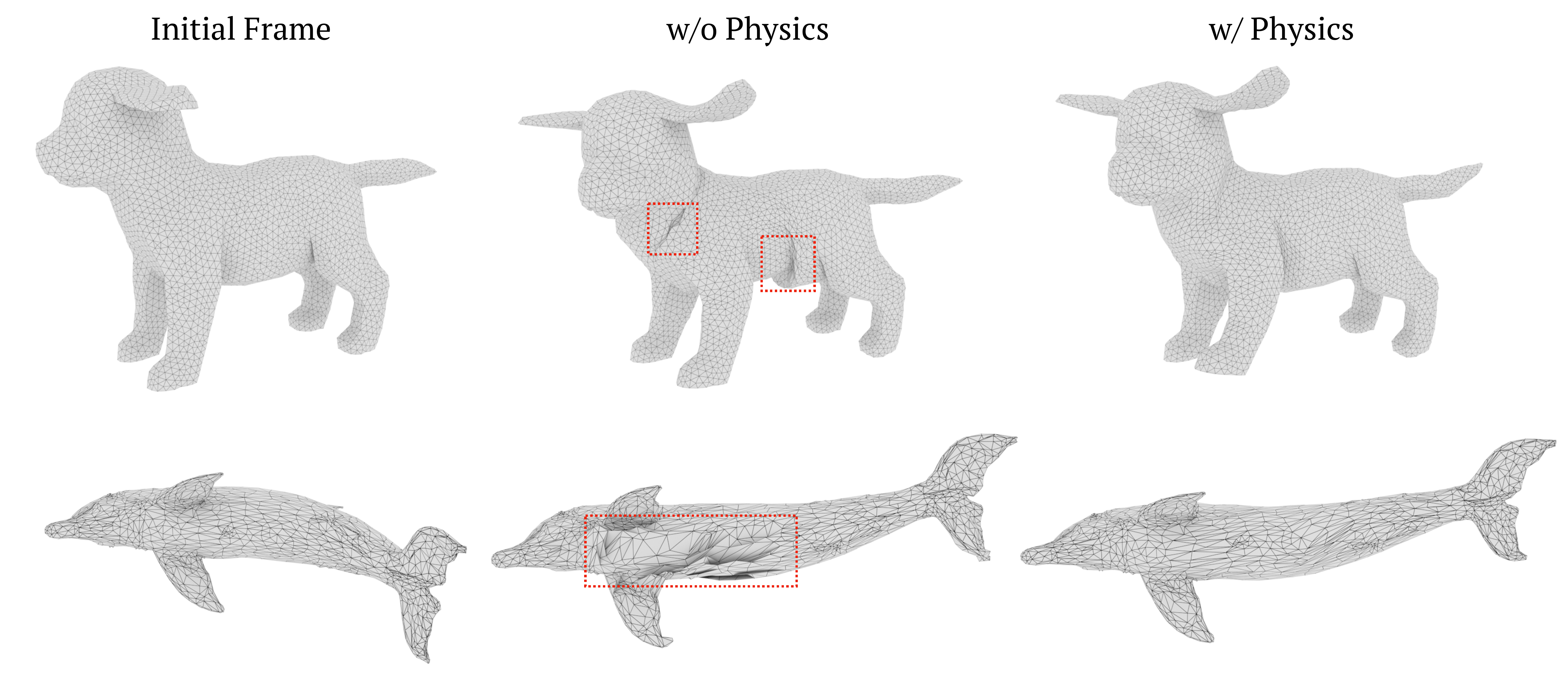}
    \caption{
        \textbf{Physics-based Refinement.} Rigged mesh surfaces often suffer from artifacts such as jittering or self-intersections (middle). Our physics-aware refinement stage eliminates such issues and enhances mesh smoothness and motion consistency.
    }
    \label{fig:ablation-physics}
    \vspace{-10pt}
\end{figure}

\paragraph{Physics-based Refinement}
We visualize mesh views of the animated results from our refinement stage in Fig.~\ref{fig:ablation-physics}. Rigging-based mesh deformation exhibits severe distortion in regions with large local deformation. Our differentiable simulation removes these surface-mesh artifacts from the rigged mesh, producing smooth geometry and motion.

% \subsection{Application}

\section{Conclusion}
We presented \METHOD, a framework for animating explicit 3D meshes using motion priors from video diffusion models. By combining differentiable rigging, video-space supervisory signals, and gradient-based pose optimization, our method transfers rich 2D motion cues into coherent 3D articulations without requiring full 4D reconstruction. A physics-aware refinement module further improves realism by enforcing temporal smoothness and physically plausible deformation. Together, these components enable controllable, editable, and simulation-ready 4D motion, offering a practical bridge between generative video models and modern animation workflows.

\newpage
{
    \small
    \bibliographystyle{ieeenat_fullname}
    \bibliography{main}
}


\newpage
\appendix
\section*{\Large Appendix}
% \clearpage
% \setcounter{page}{1}
% \maketitlesupplementary

\section{Differentiable Simulation Details} 
\paragraph{Forward Simulation}
The dynamics of deformable bodies follow Newton’s Second Law, formulated as
\begin{equation}\label{eq: motion equation}
    \frac{d^2 \bm{x}}{dt^2} = \bm{M}^{-1} \bm{f}(\bm{x})
    % (\bm{f}_{\text{int}}(\bm{x}) + \bm{f}_{\text{ext}}(\bm{x})),
\end{equation}
where $\bm{M}$ is the mass matrix, whose diagonal entries contain the masses of all vertices. The total force $\bm{f}(\bm{x})$ consists of the external forces $\bm{f}_{\text{ext}}(\bm{x})$ (e.g., gravity) and the internal elastic forces $\bm{f}_{\text{int}}(\bm{x})$ arising from deformation.
To ensure consistency with skinning-based animation, we omit external forces and simulate only the internal elastic response. The internal force $\bm{f}_{\text{int}}(\bm{x})$ captures how the deformable object attempts to restore its rest shape by resisting deformation.

In continuum mechanics, this elastic behavior is characterized by defining an energy density function $\psi(\bm{F}(\bm{x}))$, which measures the strain energy per unit undeformed volume. We employ the Fixed Corotated constitutive model \cite{stomakhin2012energetically} to describe the material behavior. Its energy density is defined as
\begin{equation}
    \psi(\bm{x}) = \mu \| \bm{F}_i - \bm{R}_i \|^2_F + \frac{\lambda}{2}(\text{det}(\bm{F}_i) - 1)^2,
\end{equation}
where $\mu$ and $\lambda$ are the Lamé parameters, $\bm{F}_i$ is the deformation gradient, and $\bm{R}_i$ is the rotational component extracted from the polar decomposition of $\bm{F}_i$. The total potential energy of the deformable body is computed as
\begin{align}
    \Psi(\bm{x}) = \sum \limits_{i=1}^N \psi(\bm{F}_i) V_i,
\end{align}
where $V_i$ denotes volume of $i$-th tetdrahedra. Note the internal resisting force is then defined as the negative gradient of the potential energy with respect to the vertex position
\begin{equation}\label{eq: internal_forces}
    \bm{f}_{\text{int}}(\bm{x}) = -\frac{\partial \Psi(\bm{x})}{\partial \bm{x}}.
\end{equation}

To advance the simulation, we use an optimization-based implicit integrator:
\begin{equation}\label{eq:optim}
    \bm{x}^{n+1} = 
    \arg\min_{\bm{x}} 
    \frac{1}{2} \|\bm{x} - \tilde{\bm{x}}\|^2_{\bm{M}} 
    + \Psi(\bm{x}),
\end{equation}
and solve this minimization via Newton’s method with line search. This yields the updated configuration $\bm{x}^{n+1}$ and allows us to compute the simulated trajectory of every vertex over time.

% The internal force fir the $\bm{f}_{\text{int}}(\bm{x})$ of Fixed Corotated deformable model can be derived as
% \begin{equation}
% \begin{aligned}
% \bm{f}_{\text{int}}(\bm{x}) & = -\frac{\partial E(\bm{x})}{\partial \bm{x}} \\
%     & = - \frac{\partial\sum\limits_{i=1}^N \Psi(\bm{F}_i) V_i}{\partial \bm{x}} \\
%     & = - \sum\limits_{i=1}^N \left( \frac{\partial \Psi(\bm{F}_i)}{\partial\bm{F}_i} \frac{\partial \bm{F}_i}{\partial \bm{x}} V_i\right), \\
% \end{aligned}
% \end{equation}

\begin{figure*}[t]
    \centering
    % --- Placeholder box for the method overview figure ---
    \includegraphics[width=\textwidth]{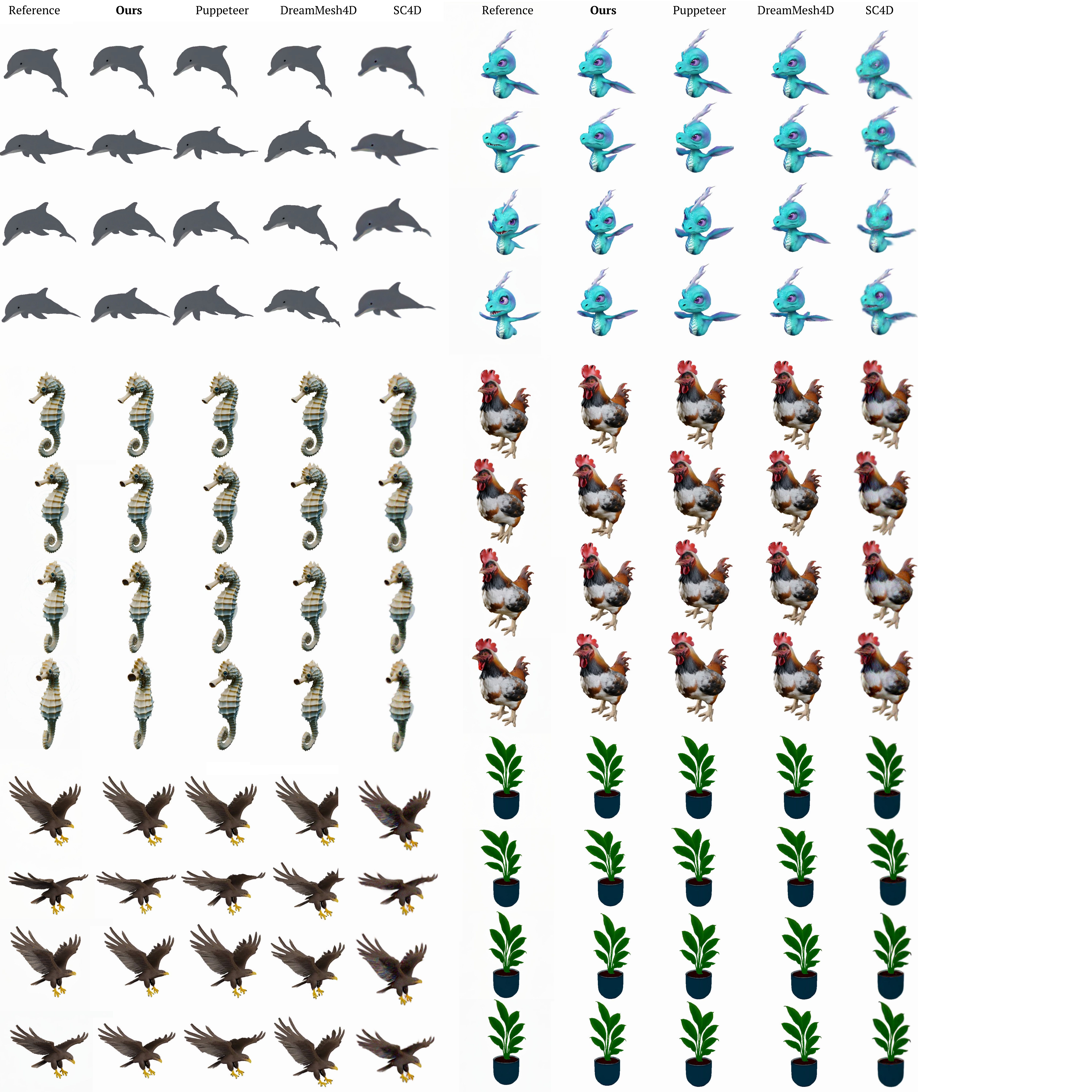}
    \caption{
    \textbf{Qualitative Comparison.} 
    We provide additional qualitative comparison results with Puppeteer~\cite{song2025puppeteer}, DreamMesh4D~\cite{li2024dreammesh4d}, and SC4D~\cite{wu2024sc4d}. %Compared to the baseline methods, our approach yields more coherent motion trajectories and more accurately reflects the dynamics present in the reference videos.
    }
    \label{fig:supp-quali-comp}
    \vspace{-10pt}
\end{figure*}

\paragraph{Backward Propagation}
Following \citet{li2025dress}, once the optimization problem in Eq.~\ref{eq:optim} is solved at each time step, we can backpropagate through the implicit integrator. For any loss function $L$, let $\bm{G}$ denote the gradient of the objective $\frac{1}{2} \|\bm{x} - \tilde{\bm{x}}\|^2_{\bm{M}} + \Psi(\bm{x}) $. The gradient can be propagated from time step $n+1$ to time step $n$ as:
\begin{align}
    \frac{dL}{d\bm{x}^n} &= -\mathcal{A}\frac{\partial \bm{G}}{\partial \bm{x}^n} - \frac{1}{h} \frac{d L}{d\bm{v}^{n+1}} \\
    \left[\frac{dL}{d\bm{v}^n}, \frac{dL}{d E} \right] &= -\mathcal{A} \left[\frac{\partial \bm{G}}{\partial \bm{v}^n}, \frac{\partial \bm{G}}{\partial E} \right],
\end{align}
where the velocity update is $\bm{v}^{n+1} = \frac{\bm{x}^{n+1} - \bm{x}^n}{\Delta t}$ and $E$ is the Young's modulus. $\mathcal{A}$ is obtained by solving a linear system:
\begin{equation}
    \mathcal{A} = \left[\frac{dL}{d\bm{x}^{n+1}} + \frac{1}{h} \frac{dL}{d\bm{v}^{n+1}} \right] \left[ \frac{\partial \bm{G}}{\partial \bm{x}^{n+1}} \right]^{-1}
\end{equation}

By combining these analytical gradients with Warp’s differentiation operator~\cite{warp2022}, our simulation becomes fully differentiable with respect to material parameters. This enables direct optimization of material properties using losses computed at arbitrary time steps.

\begin{figure}[t]
    \centering
    \includegraphics[width=\columnwidth]{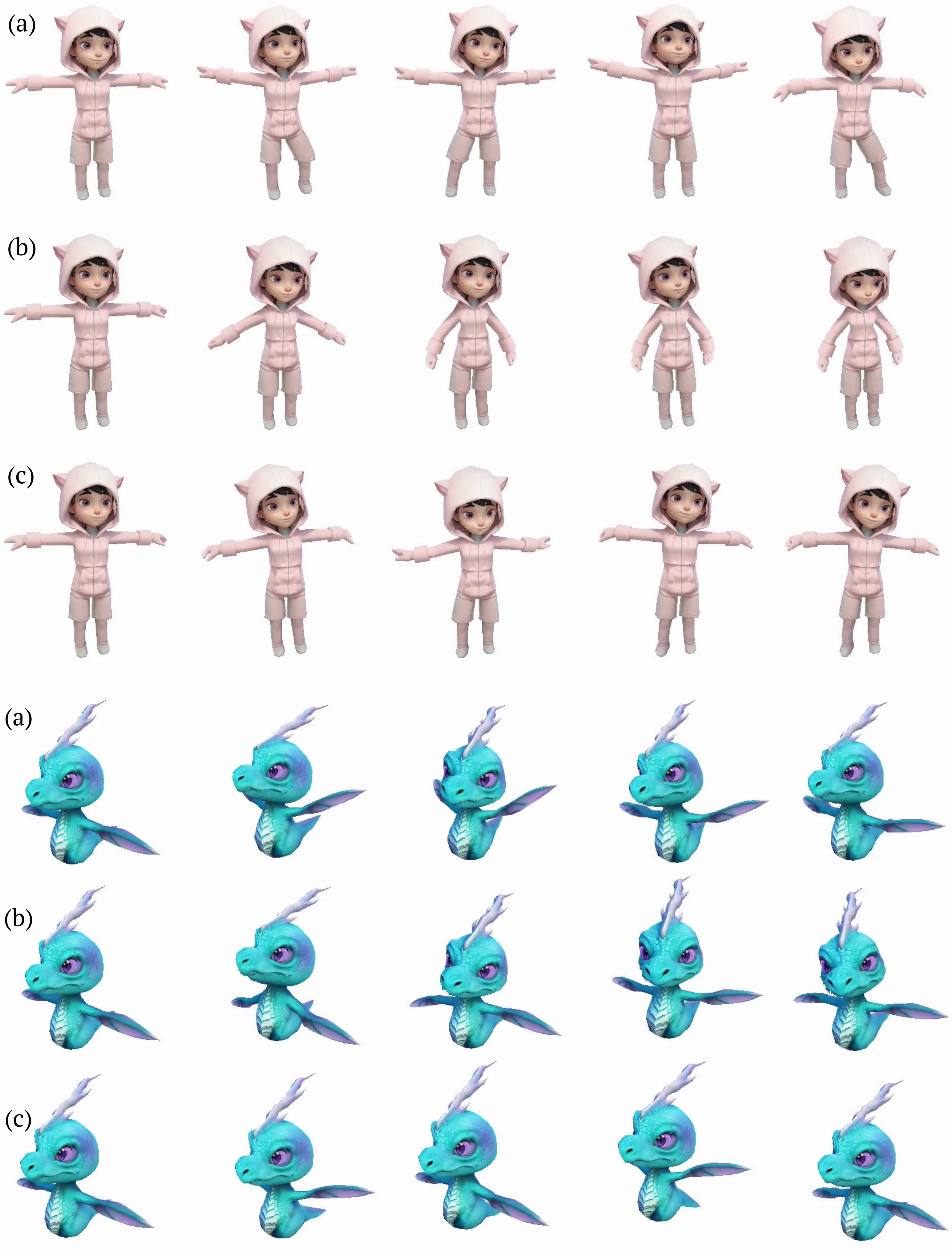}
    \caption{
        \textbf{Varying Motion Generation.} With different guiding videos of the same character, our method generates the corresponding 4D trajectories containing varied motion sequences.
    }
    \label{fig:supp-diff-motion}
    \vspace{-10pt}
\end{figure}

\begin{figure}[t]
    \centering
    \includegraphics[width=\columnwidth]{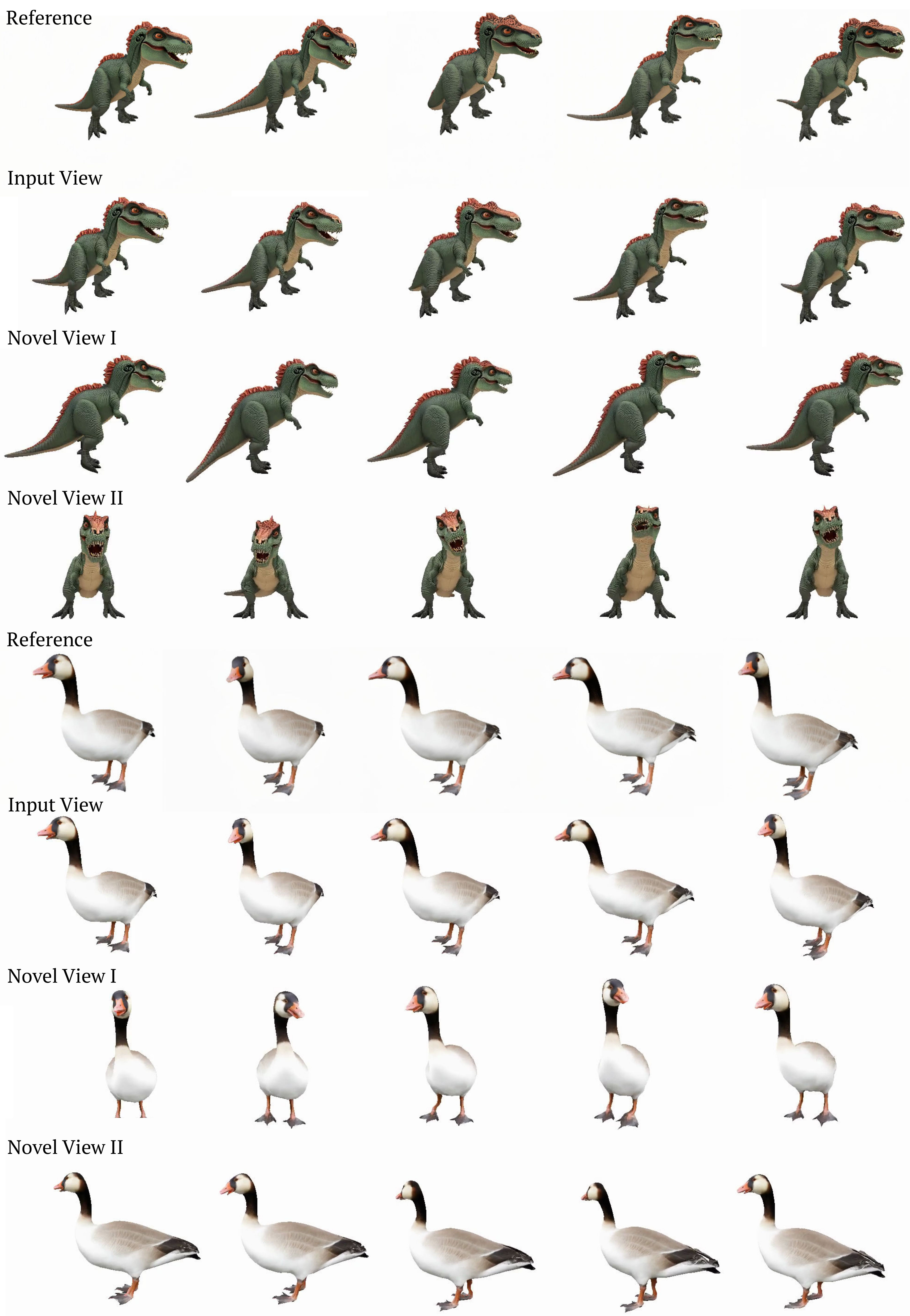}
    \caption{
        \textbf{Novel Views.} In addition to the input-view results, we render the generated animations from two novel viewpoints.
    }
    \label{fig:supp-novel-view}
    \vspace{-10pt}
\end{figure}

\section{Additional Results}

\paragraph{Additional Qualitative Comparison.}
In Fig.~\ref{fig:supp-quali-comp}, we provide additional comparison results with the baseline methods, including Puppeteer~\cite{song2025puppeteer}, DreamMesh4D~\cite{li2024dreammesh4d} and SC4D~\cite{wu2024sc4d}.

\paragraph{Diverse Motion.}
By leveraging the capability of video models, our method naturally supports generating diverse motions for the input 3D assets by producing multiple videos and optimizing the rigging model for each, as demonstrated in Fig.~\ref{fig:supp-diff-motion}.

\paragraph{Novel Views.}
We present additional novel-view results of our generated motions in Fig.~\ref{fig:supp-novel-view}. Since the motions are reconstructed on a 3D model, our method naturally maintains cross-view consistency, removing the need to generate multi-view videos.

% \section{Rationale}
% \label{sec:rationale}
% % 
% Having the supplementary compiled together with the main paper means that:
% % 
% \begin{itemize}
% \item The supplementary can back-reference sections of the main paper, for example, we can refer to \cref{sec:intro};
% \item The main paper can forward reference sub-sections within the supplementary explicitly (e.g. referring to a particular experiment); 
% \item When submitted to arXiv, the supplementary will already included at the end of the paper.
% \end{itemize}
% % 
% To split the supplementary pages from the main paper, you can use \href{https://support.apple.com/en-ca/guide/preview/prvw11793/mac#:~:text=Delete%20a%20page%20from%20a,or%20choose%20Edit%20%3E%20Delete).}{Preview (on macOS)}, \href{https://www.adobe.com/acrobat/how-to/delete-pages-from-pdf.html#:~:text=Choose%20%E2%80%9CTools%E2%80%9D%20%3E%20%E2%80%9COrganize,or%20pages%20from%20the%20file.}{Adobe Acrobat} (on all OSs), as well as \href{https://superuser.com/questions/517986/is-it-possible-to-delete-some-pages-of-a-pdf-document}{command line tools}.

\end{document}